\documentclass[11pt]{article}

\usepackage[T1]{fontenc}
\usepackage{newtxtext,newtxmath}
\usepackage[margin=1in]{geometry}
\usepackage{graphicx}
\usepackage{amsmath}
\usepackage{booktabs}
\usepackage{array}
\usepackage[hidelinks]{hyperref}
\usepackage{url}
\makeatletter
\renewcommand\normalsize{%
  \@setfontsize\normalsize{11}{13.6}%
  \abovedisplayskip 11\p@ \@plus3\p@ \@minus6\p@
  \abovedisplayshortskip \z@ \@plus3\p@
  \belowdisplayshortskip 6.5\p@ \@plus3.5\p@ \@minus3\p@
  \belowdisplayskip \abovedisplayskip
  \let\@listi\@listI}
\makeatother
\normalsize
\providecommand{\doi}[1]{\url{https://doi.org/#1}}

\title{\fontsize{18}{22}\selectfont\textbf{From Early Participation to Later Completion: Evidence from a Large-Scale Self-Paced Learning Programme}}
\author{%
\large Sakshi Sharma$^{1,*}$, Pavani Ayinampudi$^{2}$, Aditya B.M.V.$^{2}$, Jinal Gupta$^{2}$,\\
\large Prakash Hegade$^{2}$, Rohit Sharma$^{1}$, Meenakshi V$^{1}$, S.R.S. Iyengar$^{1}$\\[8pt]
\normalsize $^{1}$Indian Institute of Technology Ropar, Rupnagar, Punjab, India\\
\small\texttt{\{sakshi.23csz0006, rohit.24csz0014,}\\
\small\texttt{meenakshi.19csz0013, sudarshan\}@iitrpr.ac.in}\\[4pt]
\normalsize $^{2}$ANNAM.AI, Indian Institute of Technology Ropar, Rupnagar, Punjab, India\\
\small\texttt{\{pavania.harvard2025, adityabmv, jinalbirla, prakash.hegade\}@gmail.com}\\[6pt]
\small $^{*}$Corresponding author: \texttt{sakshi.23csz0006@iitrpr.ac.in}}
\date{}

\begin{document}
\maketitle

\begin{abstract}
Large-scale learning programmes generate records that make learner participation observable across different activities. Participation points are commonly used to record and encourage such participation, but their value may extend beyond the activities for which points are awarded. Existing evaluations often examine gamification outcomes within the activities or learning environments in which the game elements are implemented, providing limited evidence about whether early participation points contain information about later participation outside the points system. This study examines whether early participation points can provide information about learners' later participation in a self-paced learning track that does not award participation points. Using anonymised records from 876 learners in a large-scale remote software-upskilling internship, we examined participation points generated from live-session attendance and poll responses against later self-paced course completion. The primary analysis used the 438 learners who earned at least one point during the first week, while the full cohort was retained for the no-point analysis. Week-one participation points distinguished learners who later completed a self-paced course with an AUC of 0.89, increasing to 0.95 by the fourth week. Similar AUCs were observed at both stages of the self-paced course sequence, while the absence of week-one points identified learners who did not start or did not complete a self-paced course with 95\% precision. These findings indicate that early participation points can provide information about later participation outside the activities that generate the points. Such information can help large-scale learning programmes identify learners who may require timely attention while learning is still in progress, without treating participation points as a measure of overall learner engagement.
\end{abstract}

\noindent\textbf{Keywords:} Early prediction, Gamification, Learning analytics, Participation points, Participation signal, Self-paced learning

\section{Introduction}\label{sec:intro}
Participation is an important part of learning in any educational programme because it is the observable part of how learners engage with learning activities. In a face-to-face classroom, instructors can see who attends and responds, but such observation becomes difficult when hundreds of learners participate remotely, across live sessions and self-paced activities. Digital learning environments generate records of learner activity that can be analysed to understand participation and learning processes over time~\cite{siemens2011}. Such records provide information about learner participation when direct observation is difficult or not feasible at scale. The value of these records can extend beyond describing participation in individual activities when they are examined in relation to participation in other learning activities.

Gamification provides one way to structure and encourage participation in learning activities. Gamification refers to the use of game design elements in non-game contexts~\cite{deterding2011}. In educational settings, points, badges, and leaderboards are among the most commonly used game elements, particularly in e-learning environments~\cite{khaldi2023}. A learner's point total provides a simple record of participation in activities that carry points. However, the total reflects only the activities included in the point system. A high point total therefore indicates regular participation in those activities, but it does not necessarily mean that the learner participates regularly in other activities.

Learning in large-scale programmes can extend beyond the activities for which points are awarded, including synchronous sessions, self-paced courses, and assignments. A points table therefore provides one view of learner participation. The timing of recorded information also matters: a final point total describes participation across the programme but is available only after most opportunities for support have passed, whereas early participation records are available while the programme is still in progress~\cite{macfadyen2010,arnold2012}. This makes it useful to examine whether participation recorded through the points system early in the programme also relates to participation in learning activities that occur later or through other tracks, without that later participation contributing to the points total.

This study examines whether participation points earned early in a large-scale remote internship provide information about later participation in a learning activity outside the points system. The internship operated two learning tracks with separate records: a synchronous track, in which attendance and poll responses earned participation points, and a self-paced learning track, which awarded none. Because the cohort began on a common date and course completion did not contribute to the points total, week-one points can be examined against later course completion without the outcome forming part of the score. The study addresses three questions: whether the early point total is related to later completion of a self-paced course, how the information provided by participation points changes as further weeks of participation data become available, and whether the absence of points in the first week identifies learners who subsequently do not start or do not complete a course. The design is observational; the findings therefore concern predictive associations rather than causation, and participation points are not treated as a measure of learner engagement in general.

\section{Background}\label{sec:bg}
Gamification has been widely studied as an approach for supporting participation and learning in educational settings. Gamification research has examined a range of game elements, including points, badges, leaderboards, levels, and progress indicators~\cite{koivisto2019,khaldi2023}. Reviews and meta-analyses of educational gamification have reported effects on outcomes such as motivation, engagement, and academic performance, although the findings vary across learning contexts, game elements, and study designs~\cite{hamari2014,sailer2020,dichev2017}. Recent reviews have therefore given greater attention to individual game elements and to how these elements are designed and implemented rather than treating gamification as a single uniform intervention~\cite{khaldi2023,dichev2017}. Points are among the commonly used game elements in educational settings and are typically assigned when learners perform predefined activities, allowing participation to be represented through a cumulative numerical score~\cite{khaldi2023}.

Empirical studies illustrate that the effects associated with points and other game elements depend on how they are implemented and what outcome is examined. Hanus and Fox~\cite{hanus2015} examined a gamified classroom using badges and a leaderboard and reported changes in motivation, social comparison, satisfaction, effort, and academic performance over time. Mekler et al.~\cite{mekler2017} experimentally examined points, levels, and leaderboards and found effects on performance quantity, while their study did not find corresponding effects on intrinsic motivation or perceived competence in the experimental task. Adams and Du Preez~\cite{adams2022} examined the design and implementation of gamified learning activities in a university module, focusing on how gamification was incorporated into the learning activities. These studies provide evidence about how game elements operate within the learning environments in which they are introduced. They also show that the meaning and effects of a point-based system depend on the activities that generate the points and the outcomes against which the system is evaluated.

Learning analytics provides another way to examine information generated during learning. Digital learning environments produce records of learner activity that can be used to examine patterns of participation and their relationship with later outcomes~\cite{siemens2011,blumenstein2020}. Macfadyen and Dawson~\cite{macfadyen2010}, for example, used learning management system data to identify online activity variables associated with final academic performance and developed an early-warning model for educators. Kizilcec et al.~\cite{kizilcec2017} similarly examined activity patterns in massive open online courses in relation to learner strategies and goal attainment. Such work demonstrates that records generated during learning can contain information that extends beyond describing an individual activity at the time it occurs. The value of a participation record therefore depends not only on how it is constructed, but also on the outcome against which it is examined.

The timing of participation information is particularly relevant when the purpose is to identify learners who may need attention while a programme is still in progress. Macfadyen and Dawson~\cite{macfadyen2010} used early LMS activity information to develop an early-warning approach, while Arnold and Pistilli~\cite{arnold2012} described Course Signals as a learning analytics system designed to identify students who may need additional support. Temporal analysis has also been highlighted as important in learning analytics because learner activity and learning processes change over time~\cite{knight2017}. These studies establish the value of examining early participation information in relation to later outcomes. For participation points, however, an additional issue arises: the activities used to generate the points may overlap with the outcome being examined.

For participation points, the interpretation of the resulting score depends on the relationship between the activities used to generate the points and the outcome used for evaluation. When points are awarded for attendance, poll responses, or task completion, the resulting score directly records participation in those selected activities. A relationship between the points and the same activities therefore provides limited evidence about whether the information in the score extends beyond the point-generating activities. Existing gamification studies have primarily examined outcomes within the learning environments in which game elements are implemented~\cite{khaldi2023,hanus2015,mekler2017,adams2022}. Such studies are useful for understanding the role of gamification within those environments, but they do not establish whether points earned early in one set of learning activities are related to later participation in a separate learning activity that does not contribute to the point total.

The present study brings these two strands of work together by treating participation points as an early participation record and examining whether the information they contain extends to a later learning activity outside the points system. The distinction is important because the later outcome must not contribute to the original point total. When early points are generated from one learning track and later participation is recorded independently in another learning track, their relationship can be examined without making the later outcome part of the measure itself. This provides a way to examine whether an early participation record has information value beyond the activities for which points were originally awarded.

\section{Methodology}\label{sec:methods}
\subsection{Research Questions}
The study addresses one main research question:

\textbf{Main RQ.} To what extent is early participation in the synchronous track related to later participation in the self-paced learning track?

The main research question is examined through the following three sub-research questions:

\textbf{RQ1.} To what extent do early participation points provide information about later completion of a self-paced course that does not award participation points?

\textbf{RQ2.} How does the predictive performance of participation points change as more participation data become available during the programme?

\textbf{RQ3.} How effectively can the absence of participation points during the first week identify learners who subsequently do not start or do not complete a self-paced course?

\subsection{Study Setting and Cohort}
The study was conducted in a large-scale remote software-upskilling internship organised by the Vicharanashala Lab for Education Design at IIT Ropar. The programme combined instructor-led synchronous\slash live sessions and asynchronous self-paced learning activities. Learners participated in live sessions through the programme's synchronous track and could also take self-paced courses on a separate learning platform, which formed the self-paced track. The two tracks maintained separate records of learner activity.

The synchronous track used a participation-points system. Learners received points for activities such as attending live sessions and responding to in-session polls. Points were recorded in the programme's participation system and displayed through a dashboard. They provided participation feedback and recognition but did not contribute to academic grades.

The self-paced track provided courses in areas including introductory artificial intelligence and web development. Unlike the synchronous track, it did not award participation points. The platform recorded course enrolment and progress towards completion. A course was considered complete when its recorded progress reached 100\%.

The analysis used the first internship cohort to complete the programme with a common start and end date. The cohort comprised 876 learners. A common-start cohort ensured that learners had comparable calendar opportunities to participate. The two tracks were technologically separate but formed part of the same internship, allowing early participation points to be examined against a later learning outcome that did not contribute to the point total.

\subsection{Measures and Data Preparation}
Early participation in the synchronous track was measured using the cumulative participation points earned from attendance and in-session poll responses. Points were calculated directly from the underlying activity records rather than from the running leaderboard balances. To examine participation at different stages of the programme, cumulative point totals were calculated for the first week, first two weeks, first four weeks, and full observation period.

Later participation in the self-paced learning track was assessed using course progress. For this study, later participation was operationalized primarily as completion of at least one self-paced course, defined as reaching 100\% recorded progress. Course enrolment was retained as a separate measure so that learners who did not enrol could be distinguished from learners who enrolled but did not complete a course. Completion was also examined separately for the introductory artificial-intelligence course and the follow-on web-development course.

The cohort comprised 876 learners. Of these, 438 earned at least one participation point during the first week and formed the primary analysis sample, while the remaining 438, who earned no points during that week, were retained for the no-point analysis. For the early-participation analysis, week-one participation was represented both as a cumulative point total and as a binary indicator distinguishing learners with no points from those with at least one point. Week-one point tertiles and the highest point quintile were additionally used for descriptive comparisons to examine how later self-paced participation varied across different levels of early participation.

\subsection{Statistical Analysis}
Predictive performance was evaluated using the area under the receiver operating characteristic curve (AUC), which measures the ability of a continuous predictor to distinguish between two outcome groups~\cite{hanley1982}. Logistic regression was used to estimate the association between participation points and subsequent course completion. To examine how the information provided by participation points changed over time, separate models were fitted using cumulative points from the first week, first two weeks, first four weeks, and full observation period. Five-fold stratified cross-validation was used to assess the stability of the AUC estimates across folds.

Completion rates across week-one point tertiles were compared descriptively to show how later self-paced course completion varied across levels of early participation. Differences in cumulative participation points between learners who subsequently completed a self-paced course and those who did not were examined using the Mann--Whitney U test~\cite{mann1947}, with rank-biserial correlation reported as an effect-size measure~\cite{tomczak2014}. Spearman rank correlation was used to examine the association between participation points and recorded course progress~\cite{spearman1904}.

For the no-point analysis, week-one participation was represented as a binary indicator distinguishing learners who earned no participation points from those who earned at least one point. Precision and recall were calculated for identifying learners who subsequently did not start or did not complete a self-paced course. Analyses were conducted using the full observation period available for each outcome, with additional analyses applied where enrolment or completion dates were required to restrict the observation window.

\section{Data and Descriptive Statistics}\label{sec:data}
The analysis used three anonymised records: a learner list (one entry per learner, with a pseudonymous identifier and programme start and end dates), the participation-points ledger (one entry per point award, with its date and activity type), and the self-paced course record (each learner's enrolment, progress, and completion in each course). The records were joined using the pseudonymous learner identifiers. The points ledger contained 22,169 transactions, including 10,995 attendance records and 10,378 poll-response records; the remaining transactions were onboarding grants. Participation scores used in the analyses were recomputed from the underlying attendance and poll records rather than from running leaderboard balances.

The cohort contained 876 learners. Participation was unevenly distributed: the overall point total had a median of 110 and a mean of 279 points, with 43\% of learners remaining at the zero-point floor over the observation period. During the first week, 438 learners earned at least one participation point and 438 earned none.

Self-paced participation showed a separate drop-off (Table~\ref{tab:funnel}). Of the 876 learners, 460 never enrolled in a self-paced course, 416 enrolled in at least one, and 230 completed at least one course. Among the enrolled learners, 225 of 416 completed the introductory artificial-intelligence course and 163 of 216 completed the web-development course. The web-development course was the follow-on course, taken after the introductory artificial-intelligence course: all 216 learners who enrolled in it had also enrolled in the artificial-intelligence course, and 205 of them had completed it.

Among the 438 learners with at least one week-one point, 209 eventually completed at least one self-paced course and 229 did not. This active population formed the primary sample for the predictive analyses, while the full cohort was retained for reference and early-warning analyses.

\begin{table}[t]
\caption{Participation funnel for the completed cohort ($n{=}876$).}\label{tab:funnel}
\small\centering
\begin{tabular}{@{}lrr@{}}
\toprule
Stage & Learners & \% of cohort \\
\midrule
Onboarded & 876 & 100\% \\
Earned any points in week one (active) & 438 & 50\% \\
Never enrolled in any self-paced course & 460 & 52\% \\
Enrolled in $\geq$1 self-paced course & 416 & 47\% \\
\quad of whom week-one active & 351 & 40\% \\
\quad AI course enrolled / finished & 416 / 225 & 47\% / 26\% \\
\quad Web-dev course enrolled / finished & 216 / 163 & 25\% / 19\% \\
Completed $\geq$1 self-paced course & 230 & 26\% \\
\quad of whom week-one active & 209 & 24\% \\
\bottomrule
\end{tabular}
\end{table}

\section{Results}\label{sec:results}
\subsection{RQ1: Early Participation Points and Later Self-Paced Course Completion}
\begin{figure}[t]\centering
\includegraphics[width=0.62\linewidth]{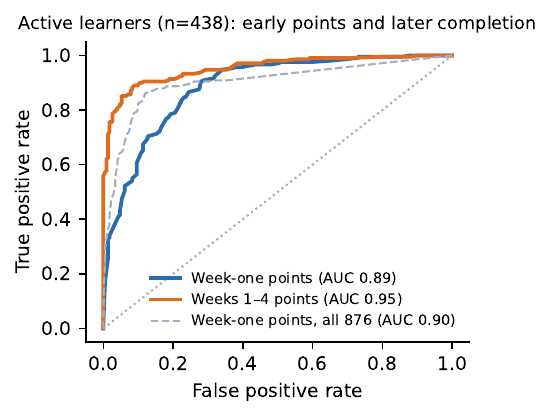}
\caption{ROC curves for predicting completion of a self-paced course from early participation points, active population ($n{=}438$). Week-one points reach AUC~0.89 and weeks 1--4 reach 0.95; the dashed curve is week-one points on the full cohort (0.90).}\label{fig:roc}
\end{figure}

The first research question examined whether early participation points were related to later completion of a self-paced course that did not award participation points. Among the 438 learners who earned at least one participation point during the first week, 209 subsequently completed at least one self-paced course, while 229 did not. Week-one participation points distinguished learners who later completed a course from those who did not, with an AUC of 0.89 (Table~\ref{tab:results}, Fig.~\ref{fig:roc}). Five-fold stratified cross-validation produced a mean AUC of 0.88 (SD $=0.05$).

A one-standard-deviation increase in week-one participation points was associated with 6.8 times higher odds of subsequently completing at least one self-paced course. Completion rates also differed across week-one point tertiles: 6\% of learners in the lowest tertile, 52\% in the middle tertile, and 85\% in the highest tertile completed at least one course.

When the two self-paced courses were examined separately, the AUC was 0.89 for the introductory artificial-intelligence course and 0.89 for the web-development course. Because the web-development course was taken after the artificial-intelligence course, these two results describe largely the same learners at two stages rather than two independent samples. Using the full cohort of 876 learners produced an AUC of 0.90. When only learners who had enrolled in at least one self-paced course were considered, the AUC was 0.81. A stricter analysis that excluded completions recorded within the predictor window and records without usable completion dates produced an AUC of 0.88 among 368 learners (Table~\ref{tab:sens}).

\begin{figure}[t]\centering
\includegraphics[width=0.62\linewidth]{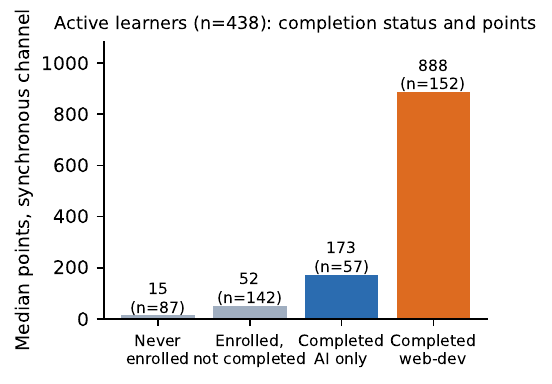}
\caption{Median cumulative participation points by self-paced course-completion status, active population ($n{=}438$). The self-paced platform awards no points.}\label{fig:ladder}
\end{figure}

The amount of participation also differed between learners who later completed a course and those who did not. Among the 438 learners with at least one week-one point, completers had a median cumulative point total of 855, compared with 29 points among non-completers ($p<10^{-60}$, rank-biserial correlation $=0.91$; Fig.~\ref{fig:ladder}). Of the 88 learners in the highest point quintile, 87 completed at least one self-paced course. The Spearman correlation between participation points and recorded course progress was 0.77 for the introductory artificial-intelligence course and 0.61 for the web-development course.

\begin{table}[t]
\caption{Key results on the active population (any week-one points, $n{=}438$), with the full cohort ($n{=}876$) for reference.}\label{tab:results}
\footnotesize\centering
\setlength{\tabcolsep}{3pt}
\begin{tabular}{@{}lrr@{}}
\toprule
 & Active & Full \\
\midrule
\multicolumn{3}{@{}l}{\textbf{Prediction of completing $\geq$1 self-paced course}}\\
Completers & 209 (48\%) & 230 (26\%) \\
AUC, week-1 points & 0.89 & 0.90 \\
AUC, weeks 1--2 / 1--4 / full period & 0.94 / 0.95 / 0.95 & 0.93 / 0.94 / 0.94 \\
5-fold CV AUC, week-1 points & $0.88{\pm}0.05$ & $0.90{\pm}0.06$ \\
AUC week-1 $\to$ AI / web-dev completion & 0.89 / 0.89 & 0.91 / 0.91 \\
Odds ratio per SD (week-1 points) & 6.8 & 7.0 \\
Completion by week-1 tertile (low / mid / high) & 6\% / 52\% / 85\% & --- \\
\midrule
\multicolumn{3}{@{}l}{\textbf{Participation of completers vs.\ non-completers}}\\
Median cumulative points, completers vs.\ not & 855 vs.\ 29 & 828 vs.\ 0 \\
Mann--Whitney $p$ / rank-biserial $r$ & $<10^{-60}$ / 0.91 & $<10^{-90}$ / 0.89 \\
Top points quintile: completed / none & 87 / 1 (of 88) & 164 / 12 (of 176) \\
Spearman $\rho$, points vs.\ progress (AI / web-dev) & 0.77 / 0.61 & 0.73 / 0.59 \\
\bottomrule
\end{tabular}
\end{table}

\begin{table}[t]
\caption{Robustness of the week-one predictor (AUC for completing $\geq$1 course) across analysis samples.}\label{tab:sens}
\footnotesize\centering
\setlength{\tabcolsep}{4pt}
\begin{tabular}{@{}lrrr@{}}
\toprule
Sample & $n$ & Completers & AUC \\
\midrule
Full cohort & 876 & 230 & 0.90 \\
Zero week-one points only & 438 & 21 (4.8\%) & 0.50 \\
Any week-one points (active) & 438 & 209 & 0.89 \\
Enrolled in $\geq$1 course & 416 & 230 & 0.81 \\
Enrolled and active & 351 & 209 & 0.84 \\
Active, excl.\ 29 completions at/before window end & 409 & 180 & 0.88 \\
Active, also excl.\ 50 undated completion flags & 368 & 139 & 0.88 \\
\bottomrule
\end{tabular}
\end{table}

\subsection{RQ2: Change in Predictive Performance Over Time}
The second research question examined how the predictive performance of participation points changed as more participation data became available during the programme. Predictive performance increased as a longer period of participation data was included. The AUC increased from 0.89 for points accumulated during the first week to 0.94 for the first two weeks and 0.95 for the first four weeks. The AUC for the full observation period was also 0.95. Thus, predictive performance reached the level observed for the full observation period by the fourth week.

Participation in the synchronous track continued after self-paced course completion for many learners. Among the 329 learners with dated course completion and participation records, only three completed a course after their final point-earning day. A total of 296 learners continued earning points for at least one week after course completion, and 261 were still earning points during the final week of the observation period. The median number of point-earning days was 52 among completers and 5 among non-completers.

\subsection{RQ3: Absence of Early Participation Points and Subsequent Non-Start or Non-Completion}
The third research question examined how effectively the absence of participation points during the first week identified learners who subsequently did not start or did not complete a self-paced course. Among the 438 learners with no week-one points, 4.8\% eventually completed at least one self-paced course, compared with 47.7\% of the 438 learners with at least one week-one point. Of the 417 learners in the no-point group who did not complete a course, 373 never enrolled in any self-paced course and 44 enrolled but did not complete one.

The week-one no-point indicator identified learners who subsequently did not start or did not complete a self-paced course with 95\% precision and 65\% recall.

\section{Discussion}\label{sec:disc}
The results show that early participation points contained substantial information about later participation in a self-paced learning track that did not contribute to the point total. Week-one participation points produced an AUC of 0.89. In this study, this means that participation during the first week of the synchronous track provided a strong distinction between learners who later completed at least one self-paced course and those who did not. Thus, information available from only the first week of participation was already sufficient to identify a substantial difference between these two groups, even though their later self-paced course activity was not included in the points system. This relationship was observed even though the points were generated from participation in the synchronous track, whereas course completion occurred in the self-paced learning track. The AUC was also 0.89 when the introductory artificial-intelligence course and the web-development course were examined separately. Because the web-development course followed the artificial-intelligence course, this reflects the same learners at two stages of the self-paced track rather than an independent replication.

The finding extends the interpretation of participation points beyond the activities from which the points are generated. In this programme, points recorded participation in attendance and in-session polls. The results show that the same early record also contained information about later participation in a separate learning activity that did not contribute to the point total. This does not mean that participation points measure learner engagement as a broad psychological construct~\cite{fredricks2004}. Rather, the findings support a narrower interpretation: early participation points can provide information about later participation in learning activities outside the activities used to generate the points.

\begin{figure}[t]\centering
\includegraphics[width=0.6\linewidth]{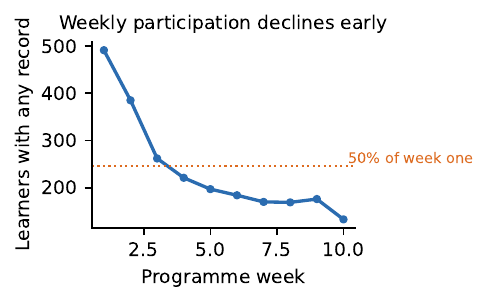}
\caption{Learners with any attendance or poll record per week. The early active population falls below half its week-one level by week four.}\label{fig:attr}
\end{figure}

The timing of this information is particularly relevant in large-scale programmes. Week-one participation points already provided substantial information about later course completion, while the AUC increased with additional participation data and reached 0.95 by the fourth week. Additional participation data after the fourth week therefore produced little further improvement in the observed AUC. This creates a practical distinction between a point total that describes participation after a programme has ended and an early participation record that provides information while the programme is still in progress. Learning analytics systems have similarly used early activity information to identify learners who may need additional support while a course is underway~\cite{macfadyen2010,arnold2012}. Participation in the synchronous track also declined quickly, as is common in online programmes~\cite{reich2019}: the number of learners with any attendance or poll record in a week, including records that earned no points, fell from 491 in the first week (438 of whom earned at least one point) to 221 in the fourth (Fig.~\ref{fig:attr}). An early indication is therefore most useful in the first week, while most learners can still be reached.

The no-point result provides a complementary view of early participation. Among learners with no participation points during the first week, only 4.8\% subsequently completed at least one self-paced course, compared with 47.7\% among learners who earned at least one point. The week-one no-point indicator identified learners who subsequently did not start or did not complete a self-paced course with 95\% precision and 65\% recall. Thus, the absence of early participation points was useful for identifying learners who subsequently did not start or did not complete a self-paced course. In a teaching or programme setting, such an early indication could help programme teams identify learners for timely follow-up, orientation, reminders, or additional support. The present study, however, does not establish whether any particular intervention would change subsequent participation or course completion.

The results also show that participation in the two learning tracks should not be understood as a simple transfer from one activity to another. Many learners who completed a self-paced course continued to earn participation points afterwards. Among learners with dated course completion and participation records, 296 of 329 continued earning points for at least one week after completing a course. Course completion was therefore not generally followed by the end of participation in the synchronous track. The association between early participation points and later course completion may instead reflect broader differences in participation across the programme. Factors such as prior interest, programme commitment, access, or other learner characteristics may contribute to participation in both tracks. Because the study is observational, the results do not establish that participation points cause later course completion.

For teachers and programme designers, the main implication is not that participation points should necessarily be introduced into every learning environment. Rather, when a programme already records participation through points or other routinely collected participation measures, those records can be examined for information about later learning activities. This is particularly relevant in large-scale and blended programmes where instructors cannot directly observe every learner across all learning activities. An early participation record from one learning track may provide useful information about participation that occurs later or through another track, before the later outcome becomes visible.

The study also suggests a useful principle for evaluating participation measures in learning systems. A measure should not be evaluated only against the activities used to construct it. When points are awarded for attendance and poll responses, for example, a relationship between the points and those same activities is partly inherent in the way the points are generated. Examining a later outcome that does not contribute to the original point total addresses a different question: whether the information captured by early participation points extends to participation in another learning activity. In this study, self-paced course completion provided such an outcome because it was not used to calculate the participation points.

This approach can be applied beyond the present internship. Large online courses, faculty-development programmes, remote internships, blended courses, and other learning environments often contain multiple learning tracks. An early participation record from one track can be examined against a later outcome from another track, provided that the later outcome does not contribute to the original measure. Such an approach can help programme teams determine whether routinely collected participation information has value beyond describing the activities in which the points were originally earned.

At the same time, the findings should be interpreted within the limits of the study. The analysis used one programme cohort and was observational. The synchronous and self-paced tracks were technologically separate but were part of the same educational programme, so the study does not establish independence between participation in the two tracks or identify the factors underlying their association. Course completion was also only one form of later participation in the self-paced track. Replication across programmes, learner populations, learning tracks, and later participation outcomes would be needed to determine how consistently early participation points provide information about later participation.

\section{Conclusion}\label{sec:conc}
This study examined whether early participation points can provide information about learners' later participation in a self-paced learning activity that does not award participation points. The findings show that participation points generated in one learning track can provide information about later participation outside the activities that generate the points. This extends the use of participation points beyond recording or encouraging the activities for which they are awarded. In large-scale learning programmes, such early information can help teachers and programme teams identify learners who may require timely attention while the programme is still in progress. However, the findings do not establish that points cause later participation or that they measure learner engagement as a broad construct. The study demonstrates a practical approach for evaluating participation measures using later outcomes that do not contribute to the original measure. Further studies across programmes and learning contexts are needed to examine how consistently this approach generalises.

\section*{Acknowledgements}
The authors thank the VicharanaShala Lab for Education Design (VLED), Indian Institute of Technology Ropar, for the opportunity to carry out this research within its internship program, for access to the anonymised operational data, and for supporting the research experiments reported here. We also thank the program's operations team and the learners whose participation made this study possible.

The authors used a generative AI assistant to help restructure and edit drafts of the text and to format the manuscript. The authors designed the study, checked all analyses, results, and references, and take full responsibility for the content.

\end{document}